\documentclass[prx,twocolumn, superscriptaddress,floatfix,longbibliography]{revtex4-2}
\usepackage{graphicx}   
\usepackage{dcolumn}    
\usepackage{bm}         
\usepackage{amsmath}    
\usepackage{amssymb}    
\usepackage{hyperref}   
\usepackage{dbnsymb} 
\usepackage{amsmath}
\usepackage{tabularx,graphicx}
\usepackage{epstopdf}
\usepackage{afterpage}
\usepackage{graphicx}
\usepackage{latexsym}
\usepackage{color, colortbl}
\usepackage{psfrag}
\usepackage{amsfonts}
\usepackage[bb=boondox]{mathalfa}
\usepackage{titlesec}
\usepackage{dsfont}
\usepackage{feynmp}
\usepackage{slashed}
\usepackage{multirow}
\usepackage{url}
\usepackage{natbib}
\usepackage{soul}
\usepackage{lipsum}

\newcommand{\hh}{{\hat{H}}}
\newcommand{\hc}{{\hat{c}}}
\newcommand{\hn}{{\hat{n}}}
\newcommand{\hg}{{\hat{g}}}
\newcommand{\bk}{{\textbf{k}}}
\newcommand{\bR}{{\textbf{R}}}
\newcommand{\ba}{{\textbf{a}}}
\newcommand{\bq}{{\textbf{q}}}
\newcommand{\bz}{{\textbf{0}}}
\newcommand{\bn}{{\textbf{m}}}
\newcommand{\Tr}{\rm{Tr}}
\newcommand{\lr}[1]{{\langle{#1}\rangle}}
\newcommand{\lnr}[1]{{\langle\hat{n}_{#1}\rangle}}

\begin{document}

	\title{
		Interplay of ferromagnetism, nematicity and Fermi surface nesting in kagome flat band
	}

    \author{Yuman He}
    \affiliation{Department of Physics, Hong Kong University of Science and Technology, Clear Water Bay, Hong Kong SAR, China}
    \affiliation{Center for Theoretical Condensed Matter Physics, Hong Kong University of Science and Technology, Clear Water Bay, Hong Kong SAR, China}

    \author{Wentao Jiang}
    \affiliation{Department of Physics, Hong Kong University of Science and Technology, Clear Water Bay, Hong Kong SAR, China}
    \affiliation{Center for Theoretical Condensed Matter Physics, Hong Kong University of Science and Technology, Clear Water Bay, Hong Kong SAR, China}

    \author{Siqi Wu}
	\affiliation{Department of Physics, Hong Kong University of Science and Technology, Clear Water Bay, Hong Kong SAR, China}
    \affiliation{Center for Theoretical Condensed Matter Physics, Hong Kong University of Science and Technology, Clear Water Bay, Hong Kong SAR, China}

	\author{Xuzhe Ying}
	\affiliation{Department of Physics, Hong Kong University of Science and Technology, Clear Water Bay, Hong Kong SAR, China}
    \affiliation{Center for Theoretical Condensed Matter Physics, Hong Kong University of Science and Technology, Clear Water Bay, Hong Kong SAR, China}

    \author{Berthold J\"ack}
	\affiliation{Department of Physics, Hong Kong University of Science and Technology, Clear Water Bay, Hong Kong SAR, China}
    
    \author{Xi Dai}
	\affiliation{Department of Physics, Hong Kong University of Science and Technology, Clear Water Bay, Hong Kong SAR, China}
    \affiliation{Center for Theoretical Condensed Matter Physics, Hong Kong University of Science and Technology, Clear Water Bay, Hong Kong SAR, China}
		
	\author{Hoi Chun Po} 
    \thanks{Corresponding author; hcpo@ust.hk}
	\affiliation{Department of Physics, Hong Kong University of Science and Technology, Clear Water Bay, Hong Kong SAR, China}
    \affiliation{Center for Theoretical Condensed Matter Physics, Hong Kong University of Science and Technology, Clear Water Bay, Hong Kong SAR, China}

\begin{abstract}
Recent experiment on Fe-doped CoSn has uncovered a series of correlated phases upon hole doping of the kagome flat bands. 
Among the phases observed, a nematic phase with a six- to two-fold rotation symmetry breaking is found to prevail over a wide doping and temperature range. Motivated by these observations, we investigate the interaction-driven phases realized in a kagome model with partially filled, weakly dispersing flat bands. Density-density interactions up to second-nearest neighbors are considered.
 We identify a close competition between ferromagnetic and nematic phases in our self-consistent Hartree–Fock calculations: while on-site interaction favors ferromagnetism, the sizable inter-sublattice interactions stabilize nematicity over a wide doping window. Competition from translational-symmetry-breaking phases is also considered.
Overall, our results show that nematicity is a generic outcome of partially filled kagome flat bands and establish a minimal framework for understanding correlated flat-band phases.
\end{abstract}

\maketitle

\section{Introduction}

The kagome lattice, constructed from corner-sharing triangles, imposes geometric frustration that induces intriguing phases of electrons in both extremes of the interaction strength: in the single-particle limit, the nearest-neighbour (NN) tight-binding (TB) model features an ideally flat band from destructive interference alongside the graphene-like dispersive bands\cite{balents2008fb_kagome}; in the strong-coupling, quantum-magnet limit, the $S=\tfrac12$ Heisenberg model on kagome serves as a paradigmatic platform for quantum spin liquids. In the intermediate-coupling regime, the kagome Hubbard model becomes a common choice to study the interplay of band-structure features and electron interactions on comparable footing. \cite{kiesel2013fRG_PRL,wang2013fRG,profe2024frg}.

Existing theoretical studies have predominantly focused on exactly flat band and dispersive bands separately, which naturally leads to two distinct origins of density of states enhancement: starting with the fully filled bands, one can first dope charge carriers into the system and a strongly correlated state is expected to emerge from the partially filled flat bands. Doping further, the flat bands can be completley depleted and when the Fermi level reaches the van Hove singularities (VHSs) in the dispersive bands, VHS-driven correlated phenomena could appear. 

In the latter regime, sublattice interference plays a crucial role, which strongly suppresses the on-site nesting channel~\cite{kiesel2012sublattice} among electronic states near the VHS, making extended interactions essential. Studies using functional renormalization group~\cite{kiesel2013fRG_PRL,wang2013fRG,profe2024frg, lco_kagome}, random phase approximation~\cite{fu2025exotic}, and variational Monte Carlo~\cite{cdw_kagome_vhs} have reported various competing instabilities including spin/charge density waves (S/CDW), superconductivity and nematic states. These reported phases are experimentally observed in paramagnetic kagome metals $\textrm{AV}_3\textrm{Sb}_5$ (A = K, Rb, Cs), where multiple dispersive-band VHSs near the Fermi level are believed to underlie the observed correlated phenomena\cite{kang2022twofold_vhs_135,vhs_cdw_135_2022,zhao2021cascade135,review2022_135,zhou2021_135_vhs_cdw,ortiz2020135,ortiz2019discover135}.

The phases which could emerge from the partial filling of the flat bands, however, are relevant to a completely different range of filling. Instead of a Fermi level close to the VHS in the dispersive band, the flat-band (FB) driven correlation is operative when the system is only slightly doped from the fully filled limit.
Rigorous ferromagnetism (FM) has been established under onsite interactions for a range of fillings near and above half-filling of the flat band~\cite{mielke1991fm_kagome_fb, tasaki2020book}, and proven to remain stable against weak dispersion at half-filling~\cite{tanaka2003stability_fm_kagome_fb}. Ref.~\cite{lin2024khm_hf} further maps out cascading magnetic textures across fillings below half-filling of the flat band with onsite repulsion, and examines extended interactions only at specific fillings.

A recent example that bridges FB and VHS physics is the kagome magnet FeGe\cite{fege_cdw_rice_2022,fege_cdw_rice_2023,yin2022fege}, which exhibits both magnetic order and charge density waves. In FeGe, the inter-layer FM is believed to originate from the partially filled kagome flat bands, while the CDW is driven by VHSs in the dispersive bands of one spin species near the Fermi level. Ref.~\cite{lin2024kgm_hf_senzhou} captures this interplay by studying the extended Hubbard model on a \emph{spin-polarized} kagome flat band, where the VHS of the minority-spin dispersive band sits near the Fermi level, naturally producing a ferromagnetic CDW phase consistent with experiment.

In contrast, the kagome paramagnet CoSn possesses fully filled flat bands and no magnetic order ~\cite{chen2024CoSn_experiment,kang2020cosn,chen2023visualizing}. Upon hole doping via Fe substitution, a series of correlated phases emerge over a wide doping and temperature range, among which a nematic phase with six- to two-fold rotational symmetry breaking is found to prevail~\cite{chen2024CoSn_experiment}. Unlike in FeGe, where FB magnetism and dispersive-band VHS instabilities cooperate, here the relevant physics is confined to the FB manifold itself. In real materials, longer-range hopping renders the nominally flat bands weakly dispersive with a bandwidth $W$ substantially smaller than the overall bandwidth, and this weak dispersion endows the flat band with its own VHS. The correlated phase diagram is then shaped by the competition between FB properties, which favor FM, and the VHS-driven instabilities within the same band, rather than a collaboration between distinct band sectors.

Starting from a simple mean-field argument, we show that inter-sublattice interactions favor nematic order in competition with the Stoner ferromagnetism driven by on-site repulsion \cite{stoner1938collective}. While inter-sublattice interactions are generally expected to be significant in kagome metals \cite{kagome_coulomb}, they are further enhanced by the extended nature of the effective FB orbitals, which are formed by superposition of orbitals along the hexagon rings of the kagome lattice \cite{balents2008fb_kagome}. Self-consistent Hartree-Fock (SCHF) calculations scanning the flat-band filling from empty to full confirm that the nematic phase persists across a sizable part of the phase diagram, in qualitative agreement with experiment \cite{chen2024CoSn_experiment}. We further show that Fermi surface nesting at the flat-band VHS can induce translational-symmetry-breaking phases; however, the heavily curved Fermi surface and weak dispersion make this enhancement less prominent and less robust against out-of-plane hopping, helping rationalize the absence of such orders in experiment.

The remainder of the paper is organised as follows. We review the non-interacting properties of the TB model with extended hopping in Section \ref{sec:tb_model}. Then the SCHF
phase diagram, band structure and the
mechanism behind the on-site and inter-sublattice interaction competition are demonstrated in Section \ref{sec:main}.

\section{Non-interacting properties}\label{sec:tb_model}
\begin{figure}
    \centering
    \includegraphics[width=1\linewidth]{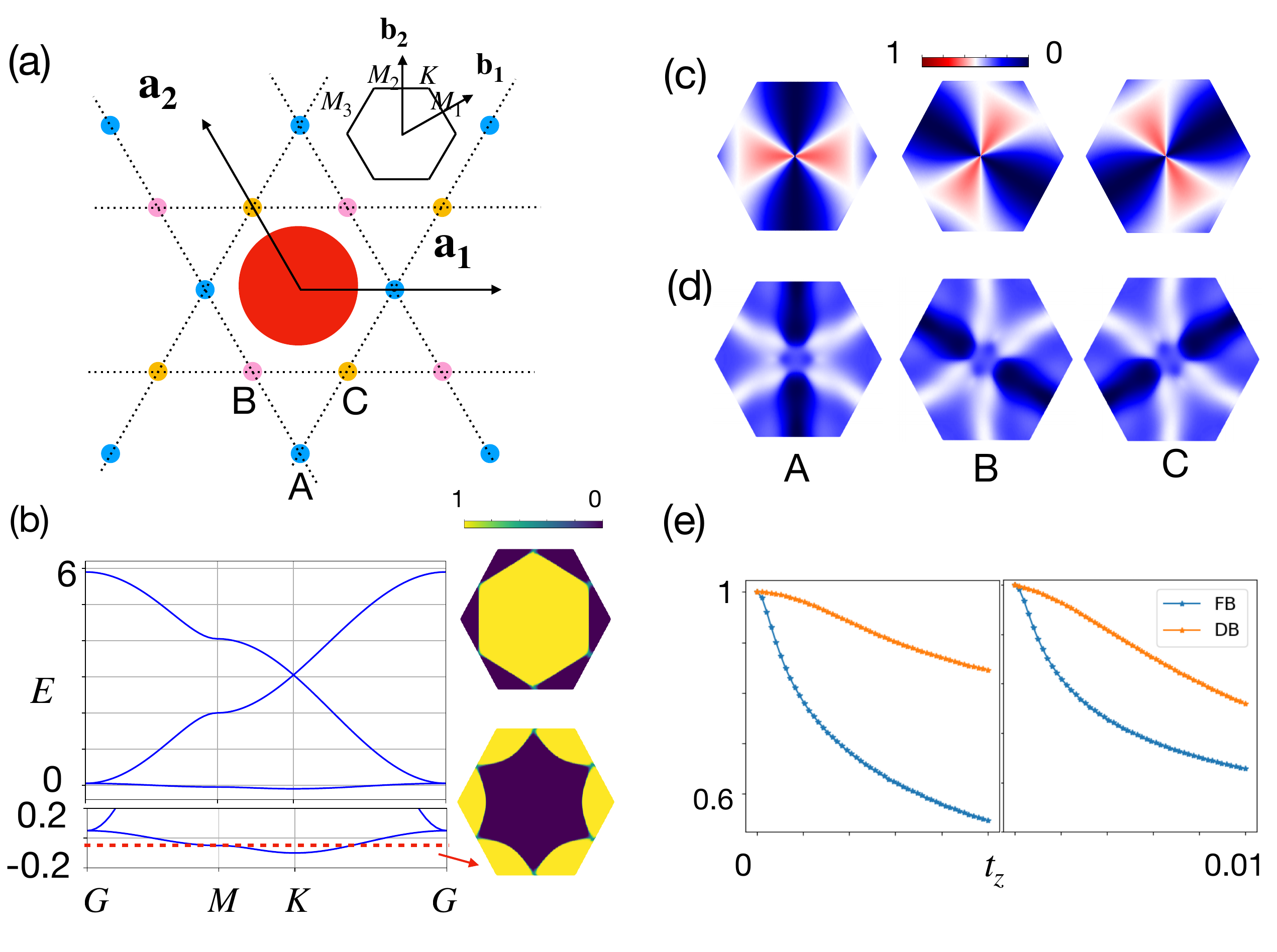}
    \caption{(a) Kagome lattice with blue, yellow and pink dots representing sites from sublattice A, B and C respectively. The red circle in the center of hexagon indicates the density distribution of hexagon states. The first Brillouin zone is attached as inset. (b) band structure of $\hh_{o}$, where $\mu=2$, $t_{1}=1$, $t_{2}=-0.025$. The upper panel shows the overall dispersion which is approximately $6t_1$. The lower panel shows the zoomed-in FB dispersion around 0.15$t_1$. Accompanying the band structure is the Fermi surface for with parallel and curved edges respectively. (c)(d) Sublattice weights over the whole BZ for toy model and realistic 48-band Wannier model of CoSn with on-site SOC\cite{chen2024CoSn_experiment} respectively, where from the left to right, the graph is for sublattice A, B and C correspondingly. (e) From left to right, $\chi_{\bz}(t_z)/\chi_{\bz}(0)$ and $\chi_{\bn}(t_z)/\chi_{\bn}(0)$ against the out-of-plane hopping $t_z$.}
    \label{fig:overall}
\end{figure}

We begin with the free-fermion TB model on the kagome lattice. Each unit cell contains three sublattices, denoted A, B, and C (shown in blue, pink and yellow in Figure~\ref{fig:overall}a). The Hamiltonian is given by:
\begin{align}\label{eq:ff_tb_model}
    \hh_{o} =&\mu\sum_{\bR i}\hc^{\dagger}_{\bR i}\hc_{\bR i} + \sum_{m\bR \delta\bR_m ij}t_{m}\hc_{\bR i}^{\dagger}\hc_{\bR+\delta\bR_m j}, 
\end{align}
where $\hat c^{\dagger}_{\mathbf{R} i}$ creates a fermion at unit cell $\mathbf{R}$ and sublattice $i$, $\mu$ is the chemical potential, $t_m$ is the $m$-th-nearest-neighbor hopping, and $\delta\bR_{m}$ runs over $m$-th-nearest-neighbor displacements. In our simulations we include hoppings up to next-nearest neighbour (NNN). Consequently, the exactly flat band generated by NN destructive interference acquires a small dispersion controlled by the ratio $t_2/t_1$. Figure~\ref{fig:overall}(c) presents the overall and corresponding FB dispersion for parameters $\mu=2$, $t_1=1$, and $t_2=-0.025t_1$, yielding a FB bandwidth of $W=0.15t_1$. 

We employ the locally rotated $d_{x^2}$-type orbitals that form the upper flat band in CoSn-family compounds\cite{kang2020cosn}. Although the flat band acquires finite dispersion, the hexagon-centered charge distribution remains the appropriate real-space configuration, as established experimentally and theoretically\cite{chen2023visualizing, kang2020cosn}. The corresponding hexagon states are non-orthonormal, which impedes SCHF convergence in metallic regimes if one implements hexagon-based density–density interactions directly. Accordingly, we adopt the conventional kagome Hubbard model with sizable inter-sublattice interaction strength induced by the extended charge distribution.

VHSs are unavoidable in two dimensions\cite{vhs_1953}. The dispersive FB exhibits a VHS at $M$ point, characterized by Hessian eigenvalues of $5t_{2}/2$ and $-3t_2/2$. The Fermi surface at van Hove filling is shown in Figure \ref{fig:overall} (b), where the upper one is for the NN TB model in the dispersive-band case, and the lower one is corresponding to the flat band in our setup. With small $|t_2/t_1|$ ratio, the Fermi surface of flat band is already heavily curved, which would result in significantly weaker nesting effects compared to the ideal case due to the saddle-point-only nesting compared to the parallel-edge nesting\cite{rice1975cdw_instability}. This effect could be quantified by computing the non-interacting Lindhard function
\begin{equation}
    \chi_{\textbf{q}}=\frac{1}{N}\sum_{\bk}\frac{f(\epsilon_{\bk+\bq})-f(\epsilon_{\bk})}{\epsilon_{\bk+\bq}-\epsilon_{\bk}},
\end{equation}
where $f(\epsilon)$ denotes the Fermi function and $\epsilon_{\vec{k}}$ represents the energy dispersion relation. The uniform response $\chi_{\bz}$(proportional to the density of states) has a logarithmic divergence at van Hove filling in regardless of the the detailed shape of Fermi surface. In contrast, the susceptibility $\chi_{\bn}$ of the nesting vector connecting $M_1$ and $M_3$, diverges only \emph{logarithmically} for the curved Fermi surface, but \emph{log-squared} for the parallel-edge one. Consequently, in the FB case the ratio $\chi_{\bn}/\chi_{\mathbf{0}}$ remains a finite constant (approximately $1.12$ in our numerical setup) rather than diverging as in the parallel-edge case. The individual $\chi_{\bn}$ and $\chi_{\mathbf{0}}$ curves are shown in the bottom panel of Figure~\ref{fig:filling_phase_diagram}(c), and their ratio is plotted versus filling in the middle panel.

We further assess the smearing of 2D VHS from out-of-plane hopping $t_z$ via the perturbation $2t_z\cos(\bk_z \ba_3)$, given experiments are conducted on CoSn thin films. As illustrated in Figure~\ref{fig:overall}(e), dispersive band, which possesses a substantially larger in-plane bandwidth, demonstrates relative robustness against finite $t_z$. Conversely, the FB case shows a marked reduction in susceptibility enhancement, suggesting weaker effective nesting in real materials.

Another notable feature of the kagome model is the non-uniform sublattice weight distribution across the first Brillouin zone (BZ), as shown in Figure \ref{fig:overall} (c). This characteristic persists even when considering real CoSn materials with spin-orbit coupling (SOC) included. (see Figure \ref{fig:overall} (d)). The non-uniform sublattice weight helps explain the resulting mean-field band structure of nematic phase and the spectral weight split observed in the experiment as we would discuss in Section \ref{sec:bs}.

\section{Phase diagrams and band structures}\label{sec:main}

In the SCHF calculations, we include interactions up to second-nearest neighbors and limit interaction strengths to $5W$, which represents a reasonably large value for flat bands and is comparable to experimentally observed gap sizes. The inter-site Coulomb interaction can be inferred from previous cRPA analyses \cite{beinevig2025kagome_1}, where since we only consider one orbital, the strength of parameters should be further reduced but the screened Coulomb interaction still contains a sizable extended component. (see discussion of cRPA in Appendix~\ref{app:crpa}.)

\subsection{Mean-field analysis}\label{sec:hf_mf}

Considering extended interactions, the interaction Hamiltonian $\hh_{I}$ could be written as:
\begin{align}
    \hh_{I} =& U\sum_{\bR i}\hn_{\bR i\uparrow}\hn_{\bR i\downarrow}\notag\\
    &+ \sum_{m\bR\delta\bR_m ij\sigma\sigma'}V_{m}\hn_{\bR i\sigma}\hn_{\bR+\delta\bR_m j\sigma'},
\end{align}
where $\hn_{\bR i\sigma}=\hc^{\dagger}_{\bR i\sigma}\hc_{\bR i\sigma}$, spin $\sigma=\uparrow\downarrow$, $U$ is the strength of on-site interaction, $V_{m}$ is the interaction strength for $m$-th-nearest neighbor. To avoid gauge complexity and focus on the ferromagnetic and nematic phases of interest, we employ collinear Hartree-Fock simulation with inter-spin correlations set to zero. The resulting mean-field Hamiltonian becomes:
\begin{align}
    \hh_{\textrm{mean}} =&U\sum_{\bR i\sigma}\lnr{\bR i\sigma}\hn_{\bR i\bar{\sigma}}\notag\\
    &+\sum_{m}V_{m}\left[\sum_{\bR i\sigma}(\sum_{\delta\bR_m,j\neq i,\sigma'}\lnr{\bR+\delta\bR_{m} j\sigma'})\hn_{\bR i\sigma}\notag \right.\\
    &\left. \qquad -\sum_{ij\sigma\delta\bR_m\bR}C_{\bR\delta\bR_{m} ij\sigma}\hat{C}_{\bR\delta\bR_{m} ij\sigma}^{\dagger} \right],
\end{align}
where constant terms have been omitted and $C_{\bR\delta\bR_{m} ij\sigma}=\lr{\hat{C}_{\bR\delta\bR_{m} ij\sigma}}=\lr{\hc_{\bR i\sigma}^{\dagger}\hc_{\bR+\delta\bR_{m} j\sigma}}$  represents the correlation function of the corresponding free-fermion state between $m$-th-nearest neighbors. In what follows, we denote the mean-field Hamiltonian as $\hh_{\textrm{mean}}[C]$ to emphasize its functional dependence on the input correlation function $C$.

Within this framework, we first verify the celebrated Stoner criterion for on-site interaction $U$ alone. In the symmetric phase, the mean-field Hamiltonian energy is $\lr{\hh_{I}}=3NUn^2$ where $N$ is the number of unit cells, $n$ is the density per spin per sublattice and $n_{\uparrow}=n_{\downarrow}=n$. A small spin imbalance $\delta n$ with new distribution $n_{\uparrow}'=n+\delta n$ and $n_{\downarrow}'=n-\delta n$, yielding an energy difference (conserving total electron number) of $\tilde{E_{I}}=3NU(n^2-\delta n^2)$. Including kinetic energy changes for spin-degenerate bands gives
 $\delta E_{K}=3N\delta n^{2}/D$, where $D=\delta n /\delta E$ is the density of states. The condition $\delta E_{I}+\delta E_{K}<0$ recovers the Stoner criterion: $UD>1$. 

Extending this analysis to include neighboring interactions, the Hartree terms yields $\delta E_{I}=\frac{3}{2}\delta n^{2}(U-4(V_1+V_2))$ and $\delta E_{K}=2N\delta n^{2}/D$ for a charge redistribution $\delta n$ from sublattice A to sublattices B and C resulting $n_{A}'=n-2\delta n$ and $n_{B}'=n_{C}'=n+\delta n$, in contrast with the original charge distribution $n_{A}=n_{B}=n_{C}=n$. This leads to a modified criterion: $(4(V_1+V_2)-U)D>\frac{4}{3}$, readily satisfied for flat bands with substantial inter-sublattice interactions. This simple analysis reveals fundamental competition between ferromagnetic and nematic phases, driven by the relative strengths of local versus inter-sublattice interactions. Since this analytical estimate neglects exchange (Fock) contributions, we proceed with full SCHF calculations for quantitative assessment.

In the SCHF procedure, we subtract the referenced-state mean-field Hamiltonian $\hh_{\textrm{mean}}[C_{o}]$ in the SCHF, where $C_{o}$ is the correlation matrix of the non-interacting state with given filling, as we focus on symmetry-breaking orders relative to the tight-binding ground state \cite{lin2024kgm_hf_senzhou,bultinck2020hf_matbg}. To identify the ground state, we employ random combinations of orthonormal basis functions constructed from charge density operators based on $D_6$ point group irreducible representations as perturbations. While our collinear SCHF framework uses charge distribution basis functions, different magnetic configurations including FM, antiferromagnetic and ferrimagnetic phases emerge through sign and magnitude differences between spin sectors; bond orders develop self-consistently during iterations. From the ensemble of converged solutions at each parameter point, we select the lowest-energy state to construct the phase diagram. We quantify symmetry breaking for operator $\hat{O}$ by $\lr{\Delta\hat{O}}=\lr{\hat{O}\hn_{i}\hat{O}^{\dagger}}- \lnr{i}$ with threshold $\delta=10^{-2}$. The details of SCHF and basis classification can be found in Appendices~\ref{app:details_SCHF} and \ref{app:seeding}, respectively.

\subsection{Translation invariant results}\label{sec:tsp_hf}
\begin{figure}
    \centering
    \includegraphics[width=\linewidth]{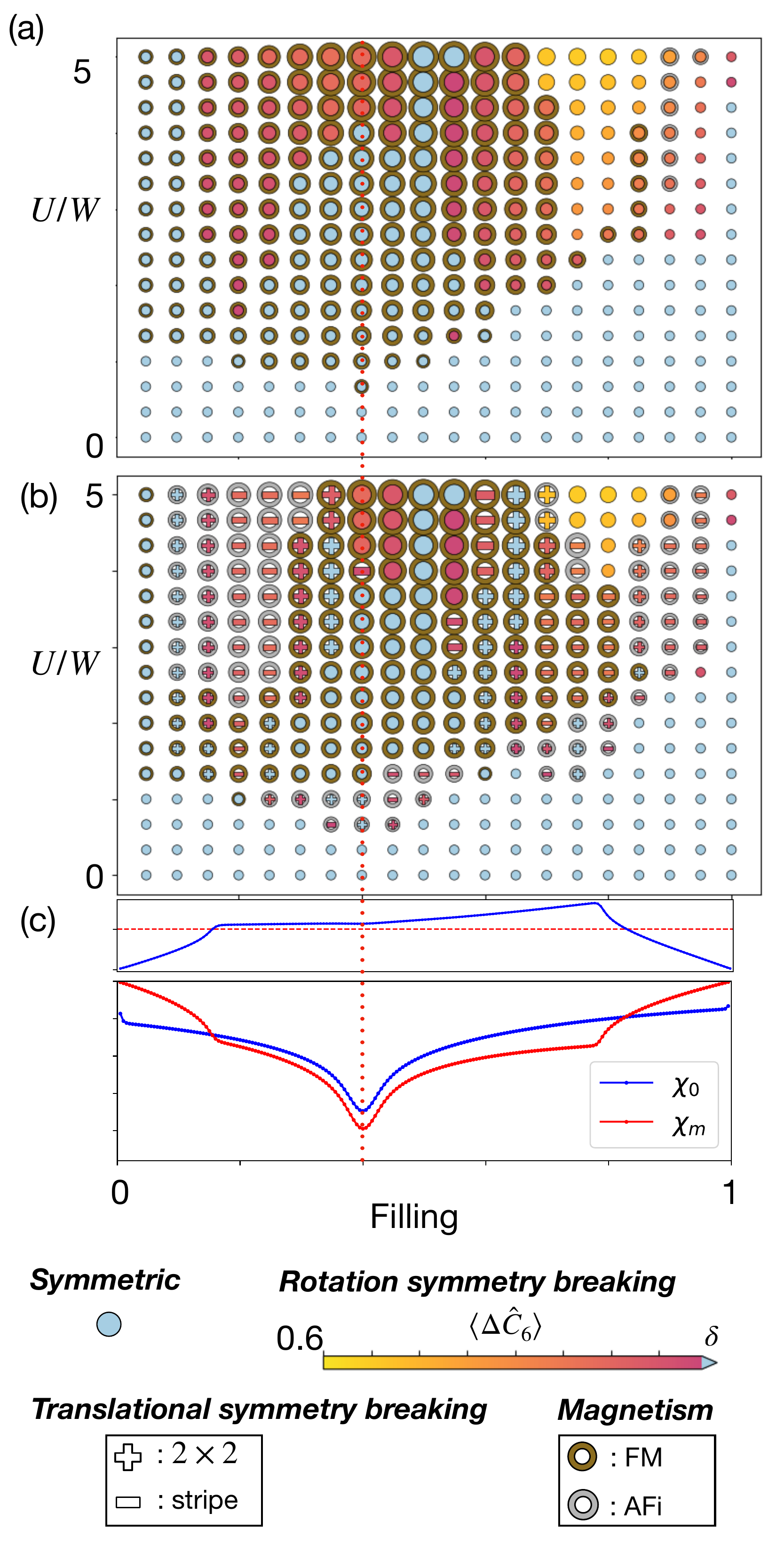}
    \caption{(a)(b) SCHF phase diagram as a function of interaction strength $U/W$ and FB filling, with fixed ratio $V_1 = V_2 = U/2$, for the translation-invariant and translation-symmetry-broken cases respectively.
A universal cutoff $\delta = 0.01$ is employed to classify symmetry breaking.
Each data point encodes three types of information.
The colormap captures the degree of nematicity through $\langle \Delta \hat{C}_6 \rangle$, with $\langle \Delta \hat{C}_6 \rangle < \delta$ marked as symmetric (blue).
The inner marker shape indicates translational symmetry: circles for translation-invariant (TI) phases, rectangles for stripe phases, and crosses for $2 \times 2$ phases; the marker size scales with the energy difference from the reference state.
The outer ring indicates magnetic properties: its thickness scales with the maximum magnetic moment $\max|\langle \hat{m}_i \rangle|$, and its color distinguishes FM (brown) from ferrimagnetic/AFM (grey) phases.
(c) The upper panel shows $\chi_\mathbf{0}/\chi_\mathbf{m}$, with the horizontal dashed red line indicating $\chi_\mathbf{0}/\chi_\mathbf{m} = 1$.
The lower panel shows the bare $\chi_\mathbf{0}$ and $\chi_\mathbf{m}$ values, both of which acquire a logarithmic divergence at filling 0.4.}
    \label{fig:filling_phase_diagram}
\end{figure}

\begin{figure}
    \centering
    \includegraphics[width=1\linewidth]{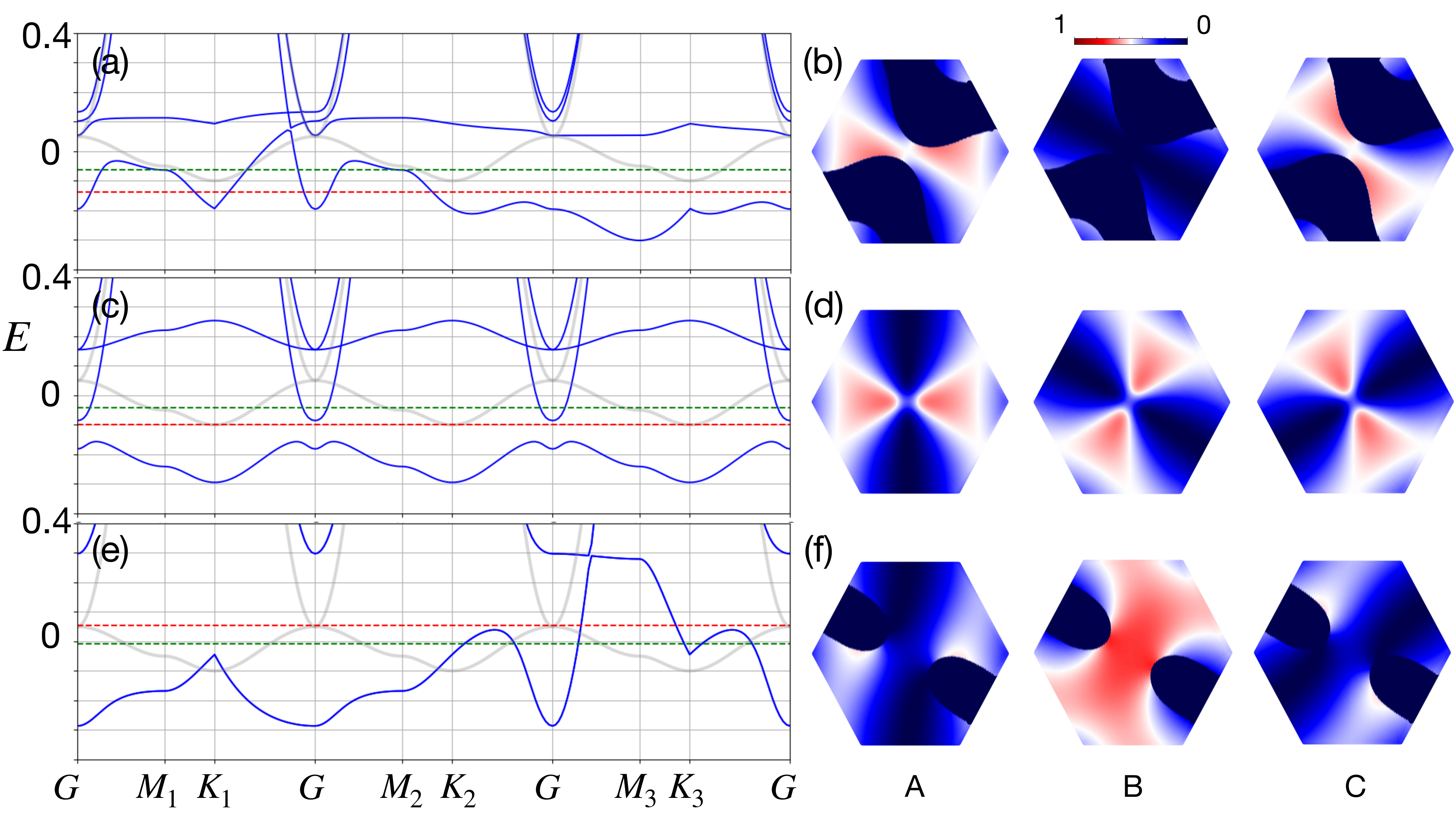}
    \caption{(a)(c)(e) Band structure for nematic FM, FM and nematic phases at $U=2V_1=2V_2=5W$ at filling 1/4, 1/2 and 3/4 respectively. The grey(blue) line is the original(SCHF) band. Dashed green(red) horizontal line is the original(SCHF) Fermi level. (b)(d)(f)Sublattice distribution per sublattice per spin over $1^{st}$ BZ for corresponding filling 1/4, 1/2 and 3/4 respectively.}
    \label{fig:bs_bz}
\end{figure}

\subsubsection{Phase diagram against filling}\label{sec:filling_pd}
We scan the entire flat band at various fillings and interaction strengths with fixed ratio $U=2V_1=2V_2$. The comprehensive phase diagram without translational symmetry breaking is shown in Figure \ref{fig:filling_phase_diagram} (a). We have classified phases from three perspectives: the breaking of six-fold rotational symmetry, which is captured by colormap; the translational symmetry breaking, which is represented by different inner marker shapes; and the magnetic properties, as the outer ring thickness indicates the maximum magnetic moment $\textrm{max}|\lr{\hat{m}_{i}}|$ with $\hat{m}_{i}=\lr{\hn_{\uparrow i}}-\lr{\hn_{\downarrow i}}$, and different ring colors differentiate FM and ferrimagnetic/AFM phases, where FM is defined as $\forall \lr{\hat{m}_{i}}>\delta$ or $\forall \lr{\hat{m}_{i}}<-\delta$. With translation invariant, all phases breaking six-fold rotation symmetry preserve 2-fold rotation symmetry. 

Overall, the system's sensitivity to small symmetry-breaking perturbations could be inferred from the minimum interaction strength required for symmetry-breaking ground states, which approximately follows the strength of $\chi_{\bz}$ at different fillings, as shown in Figure \ref{fig:filling_phase_diagram} (a)(c). The lowest symmetry-breaking threshold occurs at filling 0.4 (the VHS), while the highest threshold appears at filling 1.0 (the lower dispersive band edge), where the single FB description breaks down and interaction strengths scaled to the FB bandwidth prove insufficient for inducing strong symmetry-breaking phases.

The maximum energy difference(magnitude 0.09) occurs at half-filling with the largest interaction strength. Energy differences gradually decrease away from half-filling; similarly, they increase with interaction strength, though maximum magnetic moment $\textrm{max}|\lr{\hat{m}_{i}}|$ does not necessarily follow the same trend. In the symmetric FM case, the tendency is similar, where $\textrm{max}|\lr{\hat{m}_{i}}|$ saturates when $U\sim2W$ and remains unchanged despite continued energy difference scaling with $U$, demonstrating an energy difference peak at half-filling. However, for nematic ferromagnetic phases, $\textrm{max}|\lr{\hat{m}_{i}}|$ continues growing with interaction strength, reaching $\textrm{max}|\lr{\hat{m}_{i}}|=0.38$ at filling 0.4 and $U=5W$, which slightly exceeds the half-filled density of 0.33 due to rotational symmetry breaking.

Unlike the symmetric FM, where $\textrm{max}|\lr{\hat{m}_{i}}|$ saturates at half-filling with $|\lr{\hat{m}_{A}}=\lr{\hat{m}_{B}}=\lr{\hat{m}_{C}}|=0.33$ and decays away from this point, the nematicity energy gain from six-fold rotational symmetry breaking peaks in the upper half-filling regime. This demonstrates the capacity of sizable inter-sublattice interactions to induce strong nematicity, which generally increases with density until the single flat-band picture fails.  The largest nematicity $\lr{\Delta\hat{C}_6}=0.52$ occurs at filling 0.8 and $U=5W$. This trend explains why nematic ferromagnetic phases dominate at lower half-filling, where symmetric ferromagnetism and nematicity compete closely without either fully prevailing, particularly at large interaction scales. In the upper half-filling regime, as symmetric FM gains decay while nematicity gains continue growing, nematic phases eventually outcompete both ferromagnetic and nematic ferromagnetic phases at certain fillings.

\subsubsection{Representative band structures}\label{sec:bs}

We present representative plots for these three phases at large interaction strength( $U=2V_1=2V_2=5W$) at fillings $1/4$, $1/2$ and $3/4$ in Figure \ref{fig:bs_bz}. Figure \ref{fig:bs_bz} (c)(d) illustrate the robustness of symmetric ferromagnetic phases at half-filling. At $U\sim 2W$, the interaction strength suffices for full polarization; as interactions increase, fully polarized states become progressively gapped from remaining bands with gaps scaling with interaction strength. This is evident in the band structure of Figure \ref{fig:bs_bz} (c) and the fully occupied $1^{st}$ BZ in Figure \ref{fig:bs_bz} (d). 

Nematicity manifests through six-fold rotation related high-symmetry lines and sublattice weight distributions. In Figure \ref{fig:bs_bz} (e), the band near $M_{3}$  is substantially lifted while bands near $M_{1}$ and $M_{2}$ are lowered, corresponding to the distorted Fermi surface in Figure \ref{fig:bs_bz} (f) and nearly empty $A$, $C$ site, contrasting with the large $B$-site weight. Similar behavior occurs at $1/4$ filling, with additional full spin polarization of the lower flat band. Notably, at identical $U$, $V_1$ and $V_2$ values, nematicity is strongest at the highest filling, where sublattice B weight nearly saturates the filled Brillouin zone. This band lowering versus lifting along different rotation related momentum-space sectors also naturally explains the density of states splitting associated with nematic phases in the experiment\cite{chen2024CoSn_experiment}.

\subsubsection{Phase diagram at representative fillings}
\begin{figure}
    \centering
    \includegraphics[width=\linewidth]{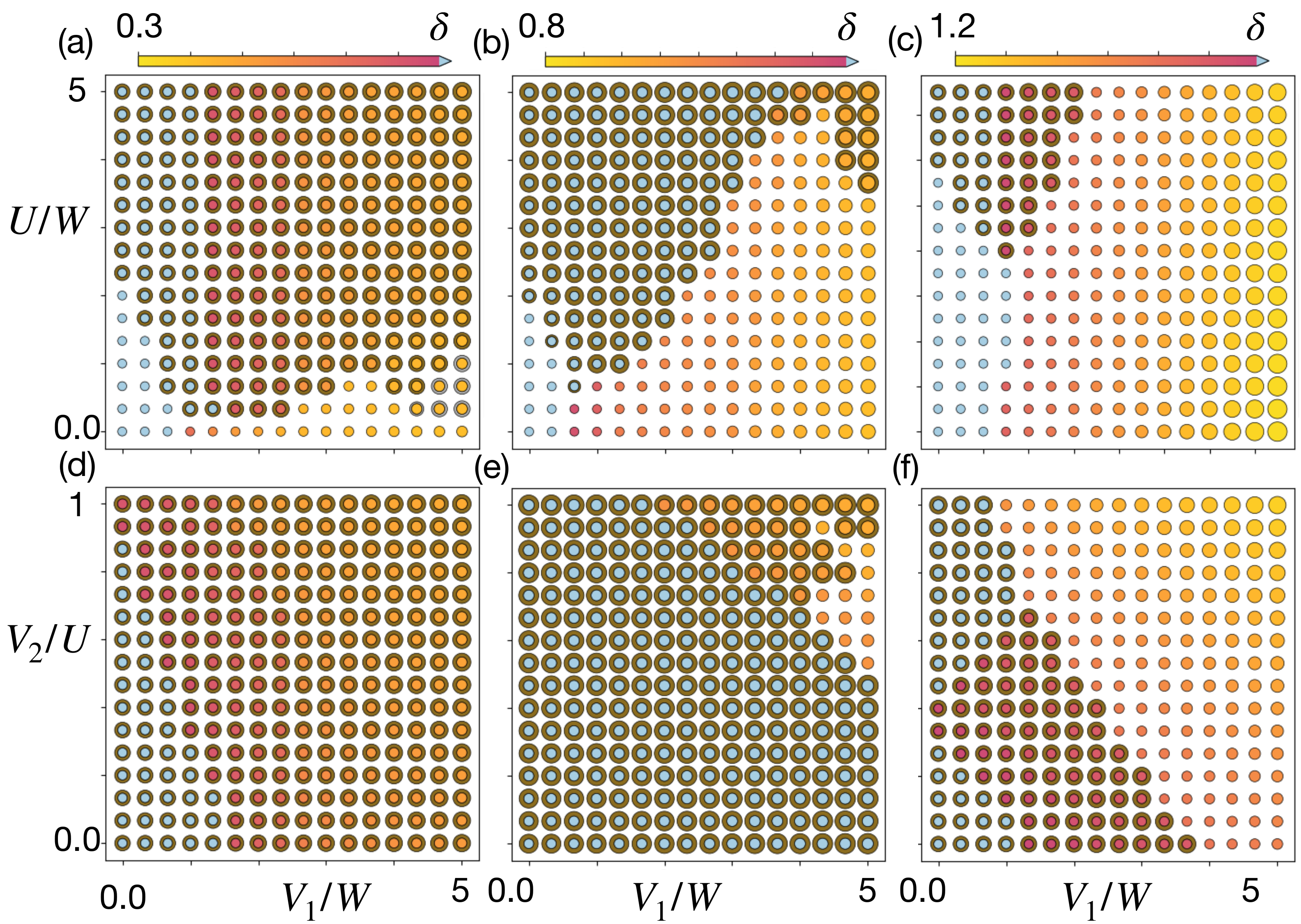}
    \caption{SCHF phase diagram at specific fillings. (a)-(c) $U$ versus $V_1=V_2$ phase diagram at filling $1/4$, $1/2$ and $3/4$. (d)-(f) $U=5W$, $V_1$ versus $V_2$ plot at filling $1/4$, $1/2$ and $3/4$. All plots share the same upper and lower bound of energy and magnetic moment scaling.}
    \label{fig:phase_diagram}
\end{figure}

While our chosen ratio follows previous cRPA analyses, we examine phase competition more specifically as functions of $U$, $V_1$ and $V_2$. Overall, increasing $V_1$ and $V_2$ enhances nematicity with almost identical driving ability if $V_1$ does not have a stronger one, as expected from analysis in Section \ref{sec:hf_mf}: both represent inter-sublattice interactions. This trend is evident in Figure \ref{fig:phase_diagram} (d-f), where we can see the FM-nematic (FM)
 phase boundaries have slopes approximately -1 or smaller than -1.

The $U$ versus $V_{1,2}$ competition visualized at different fillings reveal that nematicity could appear at a much larger $U/V$ ratio than $2$ at fractional fillings away from the half-filling, as shown in Figure \ref{fig:phase_diagram} (a) and (c). In contrast, FM is quite stable at half-filling possibly due to the gapped out, where only when $V_1+V_2$ is larger than $U$, nematicity can show up, as indicated from (b) and (e). 

Furthermore, from the overall trend of increasing filling, we can witness an enhancement of nematicity with same interaction scale, indicating the almost monotonic scaling of energy and nematicity against filling. 

\subsection{Translational-symmetry-breaking results}\label{sec:2by2_HF}
\begin{figure}
    \centering
    \includegraphics[width=\linewidth]{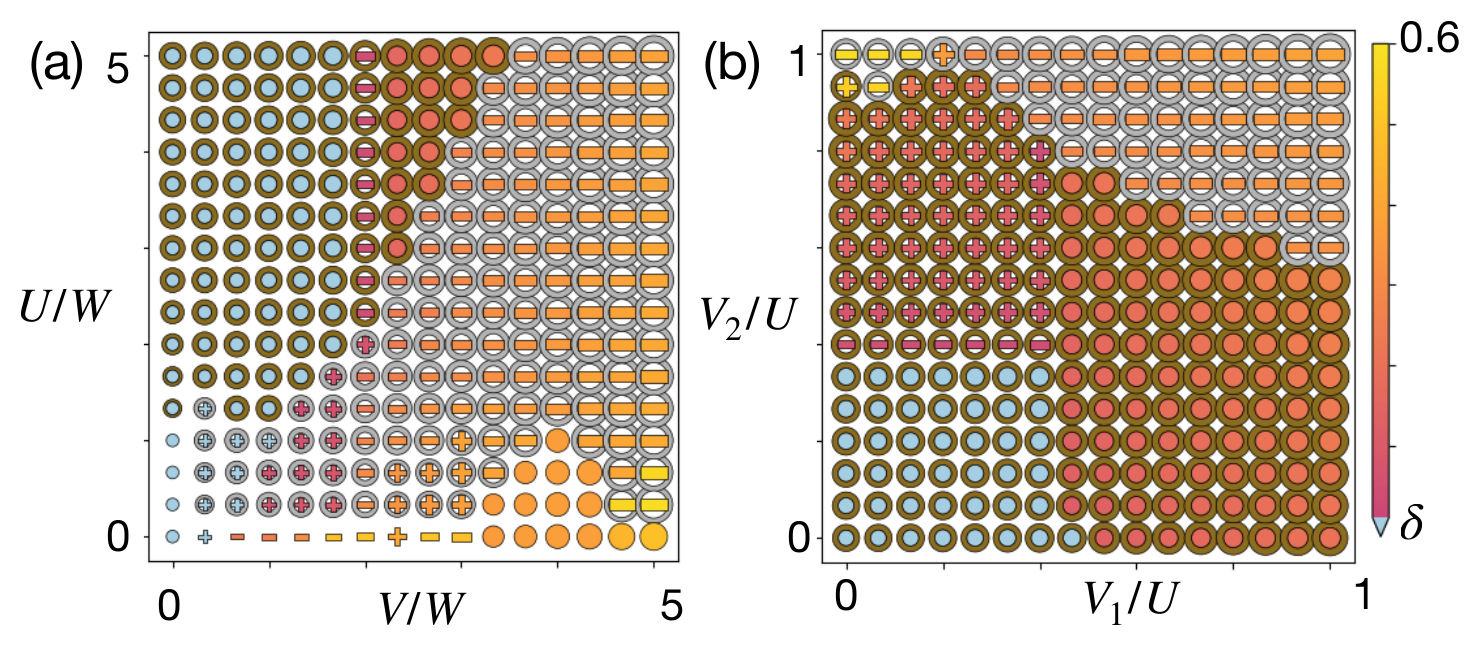}
    \caption{phase diagram for (a)$U$ versus $V_1=V_2$ and (b)$V_2$ versus $V_1$ with $U=5W$ at filling of VHS.} 
\label{fig:2_2_phase}
\end{figure}
Given the VHS at the M point in our model and experimentally observed charge density wave orders in kagome metals driven by VHSs near the Fermi level~\cite{135_cdw_2022,fege_cdw_rice_2022, fege_cdw_rice_2023,yin2022fege,135_cdw_vhs_2021,kang2022twofold_vhs_135,vhs_cdw_135_2022,zhou2021_135_vhs_cdw}, we examine how translational symmetry breaking induced by nesting could affect our phase diagram. 

We enlarge the unit cell to $2\times2$ and repeat SCHF analysis using half the $\mathbf{k}$-mesh with otherwise identical protocols. Comparing $2\times2$ and original unit-cell phase diagrams in Figure \ref{fig:filling_phase_diagram} (b) and (a), one salient distinction is the sensitivity of the system to small perturbations increases when translation breaking is taken into consideration, especially near the $\chi_{\bn}/\chi_{\bz}$ kink (indicated by upper panel in Figure \ref{fig:filling_phase_diagram} (c)), where originally symmetric phases are replaced by translational symmetry breaking phases at small interaction strength, which is consistent with the Lindhard function calculation. Similarly, ground states near van Hove filling at small interaction strength also breaks translational symmetry. However, even at the VHS and $\chi_{\bn}/\chi_{\bz}$ kink, translation-symmetry-preserving phases dominate at large interaction scales.

Apart from special fillings with absolutely and comparably large instability for translational-symmetry-breaking phases, these phases persist as ground states at other fillings even with large interactions. This implies that Fermi surface nesting might not be the driving force for them, then the possibility of artifacts cannot be completely excluded, with the mechanism behind staying unclear. 

Unlike translation-symmetry-preserving cases where all $C_6$ breaking states are $C_2$ symmetric, the doubled unit cell permits richer possibilities. Nevertheless, the colormap broadly resembles the translation-symmetry-preserving case despite some translational-symmetry-breaking phase replacements, indicating that inter-sublattice interaction driven $C_6$ breaking remains operative in doubled-unit-cell cases.

This is further evident in Figure \ref{fig:2_2_phase}, where larger $V_{1,2}$ enhances nematicity qualitatively in both plots. In Figure \ref{fig:2_2_phase} (a), we investigate a more detailed competition of $U$ versus $V_{1,2}$ at van Hove filling. At small $U$ and $V_{1,2}$, $2\times 2$ phases appear and its nematicity increases as $V_{1,2}$ grows, which is consistent with what we observe in the overall filling phase diagram. When interaction strength further grows, translationally invariant phases and stripe phases take over. More specifically, At large $U$ and intermediate $V_{1,2}$, nematic-FM phases dominate with a close competition with stripe orders.

Moreover, we investigate the phase competition between nematic FM and its neighboring phases against different $V_1$ and $V_2$ at large $U$ in Figure \ref{fig:2_2_phase} (b). Near $V_2=V_1=U/2$, close competition exists among ferromagnetic, nematic ferromagnetic, rotation-symmetry-breaking stripe, and $2\times 2$ phases. Notably, with $V_1\geq U/2$ and $V_2\leq U/2$, nematic FM phases dominate, which is consistent with results in Figure \ref{fig:2_2_phase} (a) at same $U$. At $V_2\geq U/2$ and $V_1< U/2$ region, translational-symmetry-breaking phases are favored, as $V_2/V_1>1$ would be expected to cause charge redistribution that breaks the translational symmetry. In the doubled unit cell case, we can see more distinctions between $V_1$ and $V_2$, though both are inter-sublattice interactions. When $V_1$ and $V_2$ are both larger than $U/2$, nematic stripe phases gradually take over; versus when they are both small, FM is quite robust, consistent with results in Figure \ref{fig:2_2_phase} (a).

\section{Discussions}\label{sec:discussions}
We demonstrate that the sizable inter-sublattice density-density repulsion, naturally expected from the extended nature of the charge distribution associated with the kagome flat bands, would result in nematic phases across a wide filling and interaction strength range. Considering 2D VHS at $M$ point, the possible Fermi surface nesting could induce translational-symmetry-breaking phases at fillings with (comparably) large nesting susceptibility and small interaction strength, though the nesting might be less prominent and unstable against out-of-plane hopping in thin-film real materials.

Although our analysis focuses on a single-orbital kagome model, the mechanism mainly follows from density redistribution at a given filling to reduce inter-site Coulomb energy and should therefore be insensitive to detailed orbital character, with only the quantitative phase boundaries depending on microscopic parameters.
In addition, we remark that, in the present work, we identify a sizeable inter-sublattice repulsion to be important in stabilzing the nematic phase. This could be compared with the phenomenological description we presented in Ref.\ \cite{chen2024CoSn_experiment}, which argued that when only the FB degrees of freedom are considered, the simplest order parameter is one which hybridizes both of the flat bands in CoSn, formed respectively by the in-plane and out-of-plane Co $3d$ orbitals. Bridging the SCHF analysis presented in the present work with the phenomenological description in Ref.\ \cite{chen2024CoSn_experiment} is an interesting direction for future studies.

While we do not explicitly focus on bond orders in our seeding and phase analysis, potentially relevant bond-ordered states merit future investigation. Additionally, we exclude spin-orbit coupling from our model, which might modify phase diagrams when magnetic textures become important.

\begin{acknowledgments}
YH wants to thank the helpful discussion with Nick Bultinck, Seishiro Ono, Hao Shi, Tianyu Qiao and Wangqian Miao. Y.H., W.J., X.Y., and H.C.P acknowledge support by the National Key R\&D Program of China (Grant No. 2021YFA1401500), the Hong Kong Research Grants Council (Grants No. 26308021 and C7037-22G-1), and the Croucher Foundation (Grants No. CF21SC01, CIA23SC01).
X.Y. acknowledges the support of Hong Kong Research Grants Council through Grant No. PDFS2425-6S02.
B.J. acknowledges support by the Hong Kong RGC (Grant Nos. 26304221, 16302422, 16302624, and C6033-22G) and the Croucher Foundation (Grant No. CIA22SC02). S.W. and X.D. acknowledge support by RGC General Research Fund (GRF) [Project No. 16309725]. The code is available at \cite{codes}.
\end{acknowledgments}

\bibliography{my_bib} 

@article{kiesel2012sublattice,
  title={Sublattice interference in the kagome Hubbard model},
  author={Kiesel, Maximilian L and Thomale, Ronny},
  journal={Physical Review B—Condensed Matter and Materials Physics},
  volume={86},
  number={12},
  pages={121105},
  year={2012},
  publisher={APS}
}

@article{kiesel2013fRG_PRL,
  title={Unconventional Fermi surface instabilities in the kagome Hubbard model},
  author={Kiesel, Maximilian L and Platt, Christian and Thomale, Ronny},
  journal={Physical review letters},
  volume={110},
  number={12},
  pages={126405},
  year={2013},
  publisher={APS}
}

@article{profe2024frg,
  title={Kagome Hubbard model from a functional renormalization group perspective},
  author={Profe, Jonas B and Klebl, Lennart and Grandi, Francesco and Hohmann, Hendrik and D{\"u}rrnagel, Matteo and Schwemmer, Tilman and Thomale, Ronny and Kennes, Dante M},
  journal={Physical Review Research},
  volume={6},
  number={4},
  pages={043078},
  year={2024},
  publisher={APS}
}

@article{wang2013fRG,
  title={Competing electronic orders on kagome lattices at van Hove filling},
  author={Wang, Wan-Sheng and Li, Zheng-Zhao and Xiang, Yuan-Yuan and Wang, Qiang-Hua},
  journal={Physical Review B—Condensed Matter and Materials Physics},
  volume={87},
  number={11},
  pages={115135},
  year={2013},
  publisher={APS}
}

@article{lin2024khm_hf,
  title={Complex magnetic and spatial symmetry breaking from correlations in kagome flat bands},
  author={Lin, Yu-Ping and Liu, Chunxiao and Moore, Joel E},
  journal={Physical Review B},
  volume={110},
  number={4},
  pages={L041121},
  year={2024},
  publisher={APS}
}

@article{lin2024kgm_hf_senzhou,
  title={Interplay of Charge Density Wave and Magnetism on the Kagom$\backslash$'e Lattice},
  author={Lin, Yu-Han and Dong, Jin-Wei and Fu, Ruiqing and Wu, Xian-Xin and Wang, Ziqiang and Zhou, Sen},
  journal={arXiv preprint arXiv:2409.03063},
  year={2024}
}

@article{chen2024CoSn_experiment,
  title={Cascade of strongly correlated quantum states in a partially filled kagome flat band},
  author={Chen, Caiyun and Zheng, Jiangchang and He, Yuman and Ying, Xuzhe and Sankar, Soumya and Li, Luanjing and Wei, Yizhou and Dai, Xi and Po, Hoi Chun and J{\"a}ck, Berthold},
  journal={arXiv preprint arXiv:2409.06933},
  year={2024}
}

@book{tasaki2020book,
  title={Physics and mathematics of quantum many-body systems},
  author={Tasaki, Hal},
  volume={66},
  year={2020},
  publisher={Springer}
}

@article{mielke1991fm_kagome_fb,
  title={Ferromagnetic ground states for the Hubbard model on line graphs},
  author={Mielke, Andreas},
  journal={Journal of Physics A: Mathematical and General},
  volume={24},
  number={2},
  pages={L73},
  year={1991},
  publisher={IOP Publishing}
}

@article{tanaka2003stability_fm_kagome_fb,
  title={Stability of ferromagnetism in the Hubbard model on the Kagome lattice},
  author={Tanaka, Akinori and Ueda, Hiromitsu},
  journal={Physical review letters},
  volume={90},
  number={6},
  pages={067204},
  year={2003},
  publisher={APS}
}

@article{bultinck2020hf_matbg,
  title={Ground state and hidden symmetry of magic-angle graphene at even integer filling},
  author={Bultinck, Nick and Khalaf, Eslam and Liu, Shang and Chatterjee, Shubhayu and Vishwanath, Ashvin and Zaletel, Michael P},
  journal={Physical Review X},
  volume={10},
  number={3},
  pages={031034},
  year={2020},
  publisher={APS}
}

@article{beinevig2025kagome_1,
  title={FeGe as a building block for the kagome 1: 1, 1: 6: 6, and 1: 3: 5 families: Hidden d-orbital decoupling of flat band sectors, effective models, and interaction Hamiltonians},
  author={Jiang, Yi and Hu, Haoyu and C{\u{a}}lug{\u{a}}ru, Dumitru and Felser, Claudia and Blanco-Canosa, Santiago and Weng, Hongming and Xu, Yuanfeng and Bernevig, B Andrei},
  journal={Physical Review B},
  volume={111},
  number={12},
  pages={125163},
  year={2025},
  publisher={APS}
}

@article{chen2023visualizing,
  title={Visualizing the localized electrons of a kagome flat band},
  author={Chen, Caiyun and Zheng, Jiangchang and Yu, Ruopeng and Sankar, Soumya and Law, Kam Tuen and Po, Hoi Chun and J{\"a}ck, Berthold},
  journal={Physical Review Research},
  volume={5},
  number={4},
  pages={043269},
  year={2023},
  publisher={APS}
}

@article{stoner1938collective,
  title={Collective electron ferromagnetism},
  author={Stoner, Edmund Clifton},
  journal={Proceedings of the Royal Society of London. Series A. Mathematical and Physical Sciences},
  volume={165},
  number={922},
  pages={372--414},
  year={1938},
  publisher={The Royal Society London}
}

@article{rice1975cdw_instability,
  title={New mechanism for a charge-density-wave instability},
  author={Rice, TM and Scott, GK},
  journal={Physical Review Letters},
  volume={35},
  number={2},
  pages={120},
  year={1975},
  publisher={APS}
}

@article{balents2008fb_kagome,
  title={Band touching from real-space topology in frustrated hopping models},
  author={Bergman, Doron L and Wu, Congjun and Balents, Leon},
  journal={Physical Review B—Condensed Matter and Materials Physics},
  volume={78},
  number={12},
  pages={125104},
  year={2008},
  publisher={APS}
}

@article{fege_cdw_rice_2022,
  title={Discovery of charge density wave in a kagome lattice antiferromagnet},
  author={Teng, Xiaokun and Chen, Lebing and Ye, Feng and Rosenberg, Elliott and Liu, Zhaoyu and Yin, Jia-Xin and Jiang, Yu-Xiao and Oh, Ji Seop and Hasan, M Zahid and Neubauer, Kelly J and others},
  journal={Nature},
  volume={609},
  number={7927},
  pages={490--495},
  year={2022},
  publisher={Nature Publishing Group UK London}
}

@article{fege_cdw_rice_2023,
  title={Magnetism and charge density wave order in kagome FeGe},
  author={Teng, Xiaokun and Oh, Ji Seop and Tan, Hengxin and Chen, Lebing and Huang, Jianwei and Gao, Bin and Yin, Jia-Xin and Chu, Jiun-Haw and Hashimoto, Makoto and Lu, Donghui and others},
  journal={Nature physics},
  volume={19},
  number={6},
  pages={814--822},
  year={2023},
  publisher={Nature Publishing Group UK London}
}

@article{135_cdw_2022,
  title={Rich nature of Van Hove singularities in Kagome superconductor CsV3Sb5},
  author={Hu, Yong and Wu, Xianxin and Ortiz, Brenden R and Ju, Sailong and Han, Xinloong and Ma, Junzhang and Plumb, Nicholas C and Radovic, Milan and Thomale, Ronny and Wilson, Stephen D and others},
  journal={Nature Communications},
  volume={13},
  number={1},
  pages={2220},
  year={2022},
  publisher={Nature Publishing Group UK London}
}

@article{vhs_cdw_135_2022,
  title={Charge-density-wave-induced peak-dip-hump structure and the multiband superconductivity in a kagome superconductor CsV 3 Sb 5},
  author={Lou, Rui and Fedorov, Alexander and Yin, Qiangwei and Kuibarov, Andrii and Tu, Zhijun and Gong, Chunsheng and Schwier, Eike F and B{\"u}chner, Bernd and Lei, Hechang and Borisenko, Sergey},
  journal={Physical Review Letters},
  volume={128},
  number={3},
  pages={036402},
  year={2022},
  publisher={APS}
}

@article{135_cdw_vhs_2021,
  title={Charge density waves and electronic properties of superconducting kagome metals},
  author={Tan, Hengxin and Liu, Yizhou and Wang, Ziqiang and Yan, Binghai},
  journal={Physical review letters},
  volume={127},
  number={4},
  pages={046401},
  year={2021},
  publisher={APS}
}

@article{zhou2021_135_vhs_cdw,
  title={Origin of charge density wave in the kagome metal CsV 3 Sb 5 as revealed by optical spectroscopy},
  author={Zhou, Xiaoxiang and Li, Yongkai and Fan, Xinwei and Hao, Jiahao and Dai, Yaomin and Wang, Zhiwei and Yao, Yugui and Wen, Hai-Hu},
  journal={Physical Review B},
  volume={104},
  number={4},
  pages={L041101},
  year={2021},
  publisher={APS}
}

@article{kang2022twofold_vhs_135,
  title={Twofold van Hove singularity and origin of charge order in topological kagome superconductor CsV3Sb5},
  author={Kang, Mingu and Fang, Shiang and Kim, Jeong-Kyu and Ortiz, Brenden R and Ryu, Sae Hee and Kim, Jimin and Yoo, Jonggyu and Sangiovanni, Giorgio and Di Sante, Domenico and Park, Byeong-Gyu and others},
  journal={Nature Physics},
  volume={18},
  number={3},
  pages={301--308},
  year={2022},
  publisher={Nature Publishing Group UK London}
}

@article{vhs_1953,
  title={The occurrence of singularities in the elastic frequency distribution of a crystal},
  author={Van Hove, L{\'e}on},
  journal={Physical Review},
  volume={89},
  number={6},
  pages={1189},
  year={1953},
  publisher={APS}
}

@article{kang2020cosn,
  title={Topological flat bands in frustrated kagome lattice CoSn},
  author={Kang, Mingu and Fang, Shiang and Ye, Linda and Po, Hoi Chun and Denlinger, Jonathan and Jozwiak, Chris and Bostwick, Aaron and Rotenberg, Eli and Kaxiras, Efthimios and Checkelsky, Joseph G and others},
  journal={Nature communications},
  volume={11},
  number={1},
  pages={4004},
  year={2020},
  publisher={Nature Publishing Group UK London}
}

@article{shinaoka2015accuracy,
  title={Accuracy of downfolding based on the constrained random-phase approximation},
  author={Shinaoka, Hiroshi and Troyer, Matthias and Werner, Philipp},
  journal={Physical Review B},
  volume={91},
  number={24},
  pages={245156},
  year={2015},
  publisher={APS}
}

@article{csacsiouglu2011effective,
  title={Effective Coulomb interaction in transition metals from constrained random-phase approximation},
  author={{\c{S}}a{\c{s}}{\i}o{\u{g}}lu, Ersoy and Friedrich, Christoph and Bl{\"u}gel, Stefan},
  journal={Physical Review B—Condensed Matter and Materials Physics},
  volume={83},
  number={12},
  pages={121101},
  year={2011},
  publisher={APS}
}

@article{Kresse-1993-VASP,
  title={Ab initio molecular dynamics for liquid metals},
  author={Kresse, Georg and Hafner, J{\"u}rgen},
  journal={Phys. Rev. B},
  volume={47},
  number={1},
  pages={558},
  year={1993},
  publisher={APS}
}

@article{Blochl-1994-PAW,
  title={Projector augmented-wave method},
  author={Bl{\"o}chl, Peter E},
  journal={Phys. Rev. B},
  volume={50},
  number={24},
  pages={17953},
  year={1994},
  publisher={APS}
}

@article{Perdew-1996-PBE,
  title={Generalized gradient approximation made simple},
  author={Perdew, John P and Burke, Kieron and Ernzerhof, Matthias},
  journal={Phys. Rev. Lett.},
  volume={77},
  number={18},
  pages={3865},
  year={1996},
  publisher={APS}
}

@misc{codes,
  howpublished = {\url{https://github.com/yumannhe/SCHF_auto_collinear}}
}

@article{kagome_coulomb,
  title={Electronic correlations and universal long-range scaling in kagome metals},
  author={Di Sante, Domenico and Kim, Bongjae and Hanke, Werner and Wehling, Tim and Franchini, Cesare and Thomale, Ronny and Sangiovanni, Giorgio},
  journal={Physical Review Research},
  volume={5},
  number={1},
  pages={L012008},
  year={2023},
  publisher={APS}
}

@article{cdw_kagome_vhs,
  title={Charge density waves in kagome-lattice extended Hubbard models at the van Hove filling},
  author={Ferrari, Francesco and Becca, Federico and Valent{\'\i}, Roser},
  journal={Physical Review B},
  volume={106},
  number={8},
  pages={L081107},
  year={2022},
  publisher={APS}
}

@article{lco_kagome,
  title={Loop current order on the kagome lattice},
  author={Zhan, Jun and Hohmann, Hendrik and D{\"u}rrnagel, Matteo and Fu, Ruiqing and Zhou, Sen and Wang, Ziqiang and Thomale, Ronny and Wu, Xianxin and Hu, Jiangping},
  journal={arXiv preprint arXiv:2506.01648},
  year={2025}
}

@article{fu2025exotic,
  title={Exotic charge-density waves and superconductivity on the kagome lattice},
  author={Fu, Ruiqing and Zhan, Jun and D{\"u}rrnagel, Matteo and Hohmann, Hendrik and Thomale, Ronny and Hu, Jiangping and Wang, Ziqiang and Zhou, Sen and Wu, Xianxin},
  journal={National Science Review},
  volume={12},
  number={11},
  pages={nwaf414},
  year={2025},
  publisher={Oxford University Press}
}

@article{zhao2021cascade135,
  title={Cascade of correlated electron states in the kagome superconductor CsV3Sb5},
  author={Zhao, He and Li, Hong and Ortiz, Brenden R and Teicher, Samuel ML and Park, Takamori and Ye, Mengxing and Wang, Ziqiang and Balents, Leon and Wilson, Stephen D and Zeljkovic, Ilija},
  journal={Nature},
  volume={599},
  number={7884},
  pages={216--221},
  year={2021},
  publisher={Nature Publishing Group UK London}
}

@article{review2022_135,
  title={Charge order and superconductivity in kagome materials},
  author={Neupert, Titus and Denner, M Michael and Yin, Jia-Xin and Thomale, Ronny and Hasan, M Zahid},
  journal={Nature Physics},
  volume={18},
  number={2},
  pages={137--143},
  year={2022},
  publisher={Nature Publishing Group UK London}
}

@article{yin2022fege,
  title={Discovery of charge order and corresponding edge state in kagome magnet FeGe},
  author={Yin, Jia-Xin and Jiang, Yu-Xiao and Teng, Xiaokun and Hossain, Md Shafayat and Mardanya, Sougata and Chang, Tay-Rong and Ye, Zijin and Xu, Gang and Denner, M Michael and Neupert, Titus and others},
  journal={Physical review letters},
  volume={129},
  number={16},
  pages={166401},
  year={2022},
  publisher={APS}
}

@article{ortiz2019discover135,
  title={New kagome prototype materials: discovery of KV 3 Sb 5, RbV 3 Sb 5, and CsV 3 Sb 5},
  author={Ortiz, Brenden R and Gomes, L{\'\i}dia C and Morey, Jennifer R and Winiarski, Michal and Bordelon, Mitchell and Mangum, John S and Oswald, Iain WH and Rodriguez-Rivera, Jose A and Neilson, James R and Wilson, Stephen D and others},
  journal={Physical Review Materials},
  volume={3},
  number={9},
  pages={094407},
  year={2019},
  publisher={APS}
}

@article{ortiz2020135,
  title={Cs V 3 Sb 5: AZ 2 topological kagome metal with a superconducting ground state},
  author={Ortiz, Brenden R and Teicher, Samuel ML and Hu, Yong and Zuo, Julia L and Sarte, Paul M and Schueller, Emily C and Abeykoon, AM Milinda and Krogstad, Matthew J and Rosenkranz, Stephan and Osborn, Raymond and others},
  journal={Physical Review Letters},
  volume={125},
  number={24},
  pages={247002},
  year={2020},
  publisher={APS}
}

\appendix

\section{SCHF}\label{app:details_SCHF}

In the main text we only mention the real-space mean-field decomposition, while because of the translational symmetry manifestation, we Fourier transform the operators into momentum space, and stick to the periodic gauge:
\begin{align}
    \hc_{\bk\sigma}^{\dagger} = \frac{1}{\sqrt{N}}\sum_{\bR}\hc_{\bR \alpha\sigma}^{\dagger}e^{-i\bk \bR},
\end{align}
Then the TB Hamiltonian $\hh_{o}$ becomes:
\begin{align}
    \hh_{o} =& \sum_{\bk,\alpha}\mu\hc_{\bk,\alpha}^{\dagger}\hc_{\bk,\alpha}+\notag\\&\sum_{\bk,i}(t_i\sum_{\alpha\beta\delta\bR_i}e^{-i\bk\delta\bR_i})\hc_{\bk\alpha}^{\dagger}\hc_{\bk\beta},
\end{align}
where $\delta\bR_i$ is the unit-cell displacement for $i^{th}$ nearest neighbor hopping, which takes values of $n\hat{a}_1+m\hat{a}_2$ with $n$ and $m$ integers. The mean-field Hamiltonian could also be rewritten in momentum space. Since we restrict the system to be collinear, the mean-field decomposition becomes:
\begin{align}
    \hh_{\textrm{mean}}=&U\sum_{\bk,\alpha,\sigma}n_{\alpha\sigma}\hn_{\bk\alpha\bar{\sigma}}+\notag\\
    &2(V_1+V_2)\sum_{\bk\alpha\sigma}(\sum_{\beta\neq\alpha,\sigma'}n_{\beta\sigma'})\hn_{\bk\alpha\sigma}-\notag\\
    &\sum_{i=1}^{i=2}(V_i\sum_{\bk,\alpha,\beta,\delta\bR_i,\sigma}C_{\delta\bR_i\alpha\beta\sigma}\hat{C}_{\bk\alpha\beta\bar{\sigma}}^{\dagger}e^{i\bk\delta\bR_i}).
\end{align}
Since we are interested in symmetry-breaking orders, we take the TB model ground state as the reference state, as symmetric interaction has been considered, and we have already controlled the flat band bandwidth similar to the DFT result during this procedure. Therefore, for $i^{th}$ iteration of the SCHF, we diagonalize $\hh_{i}=\hh_{o} + \hh_{\textrm{mean}}[C_{i-1}]- \hh_{\textrm{mean}}[C_{o}]$ in $\bk$-space for each $\bk$. $C_{i-1}$($C_{o}$) is the correlation function from $i^{th}$ iteration (of TB ground state). To avoid double counting, the corresponding energy is calculated as: $E_{i}=\Tr[(H_{o}+H_{\textrm{mean}}[C_{i-1}]/2-H_{\textrm{mean}}[C_{o}])^{T}C_{i}]+\Tr[H_{\textrm{mean}}^{T}[C_{o}]C_{o}]/2$. We can check that when the input state $C_{i-1}=C_{o}$, $\hh_{i}=\hh_{o}$ and $E_{i}=\Tr[H_{o}^{T}C_{o}]=\sum_{ij}H_{o,ij}C_{o,ij}=\sum_{ij}H_{o,ij}\lr{\hc_{i}^{\dagger}\hc_{j}}_{o}=\lr{\hh_{o}}$, which is the ground state energy without interaction. 

A subtle part of the partially filled flat-band system is the finite-size effect, where the large degeneracy at Fermi level should be treated carefully to avoid the mismatch in electron count. This problem is especially crucial when we consider the filling exactly at the VHS, where small difference in filling could result in different state. Consequently, we add pseudo-temperature to smoothen the Fermi surface. (Details of tests are in Appendix. \ref{app:t_test}) For the grid $60\times60$, we use $k_{B}T=4\times10^{-4}$, with Fermi-Dirac distribution, the altered occupation for $i^{th}$ eigenvalue $\epsilon_{\bk i\sigma}$ of spin $\sigma$ at $\bk$ with Fermi level $\mu$ is $\tilde{n}_{\bk i\sigma}=\frac{1}{e^{(\epsilon_{\bk i\sigma}-\mu)/k_{B}T}}$. The corresponding $\lr{\hat{C}_{\bk mn \sigma}}=\sum_{i}v_{\bk ni\sigma}\tilde{n}_{\bk i\sigma}v_{\bk mi\sigma}^{*}$, where $\hat{v}_{\bk i\sigma}^{\dagger}$ is the corresponding eigenmode for energy $\epsilon_{\bk i\sigma}$, with $\hat{v}_{\bk i\sigma}^{\dagger} = \sum_{m}\hc_{\bk m\sigma}^{\dagger}v_{\bk mi\sigma}$. Therefore, $\hc_{\bk m\sigma}^{\dagger} = \sum_{i}\hat{v}_{\bk i\sigma}^{\dagger}v_{\bk im\sigma}^{*}$, then $\hc_{\bk m\sigma}^{\dagger}\hc_{\bk n\sigma}=\sum_{ij}\hat{v}_{\bk i\sigma}^{\dagger}v_{\bk im\sigma}^{*}\hat{v}_{\bk j\sigma}v_{\bk nj\sigma}$. Since eigenmodes are orthonormal except for the occupation number is smoothened by pseudo-temperature, therefore $\lr{\hat{C}_{\bk mn \sigma}}=\sum_{i}v_{\bk ni\sigma}\tilde{n}_{\bk i\sigma}v_{\bk mi\sigma}^{*}$.

since $\hat{C}_{\delta\bR\alpha\beta\sigma} = \sum_{\bk}\hat{C}_{\bk\alpha\beta\sigma}e^{-i\bk\delta\bR}$, in this way we can construct $\hh_{\textrm{mean}}$ for next iteration based on the correlation function.

For the input initial seed of $C_{0}$, we employ random seeds with small strength, which acts as small perturbation to the symmetric states, and see if the symmetry-breaking orders could develop. (See Appendix. \ref{app:seeding} for details of classification of basis)

We set the convergence criteria to be $C_{i}-C_{i-1}<10^{-7}$, where most translational symmetry preserving seeds can meet the criteria, even in the worst case, $C_{f}-C_{f-1}<10^{-4}$ with maximum iteration set to be $f=500$. In translational symmetry breaking cases, more seeds would not converge at all. Therefore we exclude all results without convergence satisfying $C_{i}-C_{i-1}>10^{-4}$. If the phase diagram is not smooth enough, we would try more runs of random generation until the diagram is smooth enough.

\section{pseudo-temperature test}\label{app:t_test}
\begin{figure}
    \centering
    \includegraphics[width=\linewidth]{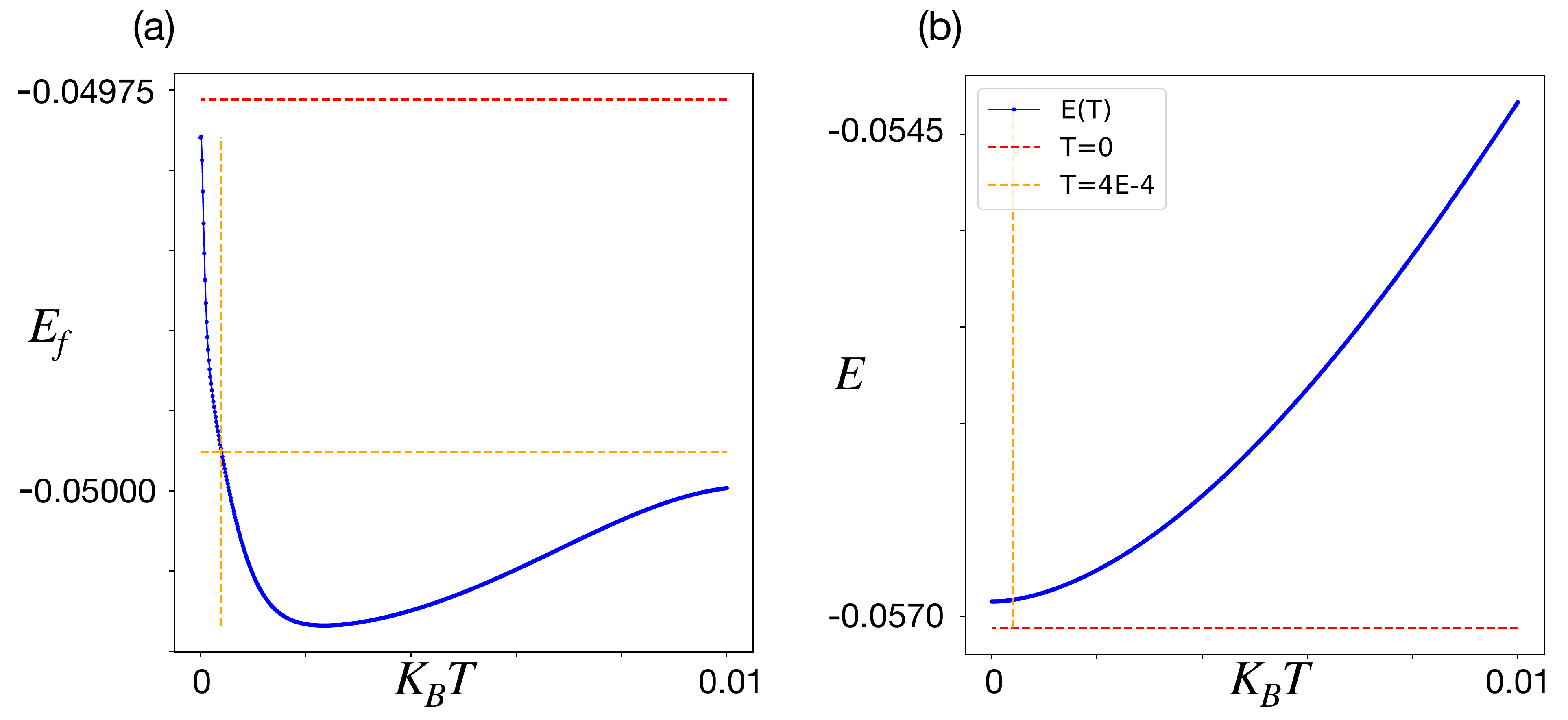}
    \caption{(a)(b) Fermi level and Ground state energy change with respect to pseudo temperature respectively. The red and yellow dashed line corresponds to zero temperature (direct count) and the temperature we used in the simulation accordingly.}
    \label{fig:temp_test}
\end{figure}

In this section we show the temperature test for the case when the filling is exactly at the VHS. Without smoothening, the large degeneracy at the Fermi level would cause the direct counting to have an occupation of 0.401 per unit cell, while the exact filling is 0.4. After adding pseudo-temperature for smoothening the Fermi level, we can get the exact filling, while we test the ground state energy and Fermi level change with temperature plotted in Figure \ref{fig:temp_test}. We can see that the miscounting would also result in discrepancies in ground state energy and Fermi energy, which is indicated by red line, while the orange line indicating the temperature we used in the SCHF, which gives much better resolution than the bare count. The exact Fermi energy should be at -0.05. 

\section{Seeding states classification}\label{app:seeding}
Based on the $D_6$ point group and translational symmetry, we can classify spatial states into irreducible representations (irrep). For 1D irrep, the case is straightforward where for given symmetry operator $\hg$, character $\chi$, we have basis $\Delta$ that satisfies:
\begin{align}
    \hg\Delta=\chi\Delta,\notag\\
    (\hg-\chi)\Delta=0.
\end{align}
Considering all m generators and corresponding characters for given irrep, we have the equation:
\begin{align}\label{eq:seeding_symm}
    \begin{bmatrix}
        \hg_{1}-\chi_1\hat{\mathbb{1}}\\
        \vdots\\
        \hg_{m}-\chi_m\hat{\mathbb{1}}
    \end{bmatrix}\cdot \Delta=G\cdot\Delta=0.
\end{align}
To solve eq. \ref{eq:seeding_symm}, we apply singular value decomposition (SVD) to $G$, where $G = U\Gamma V^{\dagger}$, with $U$, $V$ isometries and $\Gamma$ diagonal matrices with singular values the entries. Then naturally, the null vectors $V_i$ in $V$ corresponding to 0 singular values are the solutions to the equation $G\cdot\Delta=0$, as:
\begin{align}
    GV_i = U\cdot 0=0.
\end{align}

Similarly, for higher-dimensional irreps, we have:
\begin{align}
    \hat{g}_{ij}\Delta_{jm} = \Delta_{il}u_{lm}\\
    (g_{ij}\delta_{lm} - \delta_{ij}u_{lm})\delta_{jl} = 0.
\end{align}
where $u$ is the transformation matrix for the 2D irrep basis. Then similarly the basis for higher-dimensional irreps are the null vectors for the operator $(g_{ij}\delta_{lm} - \delta_{ij}u_{lm})$.

\section{cRPA estimate on interaction strengths}\label{app:crpa}
The topological nature of the flat bands prevents the direct use of standard constrained Random Phase Approximation (cRPA) packages, such as VASP, to evaluate the screened interaction parameters. One proposed approach involves using the compact localized states associated with the d-orbital flat band as a trial basis for constructing Wannier functions\cite{kang2020cosn}.
\begin{align}
    U^{\prime ij}_{kl} = U^{ij}_{mn}\left[ \left(I + 2(\chi - \tilde{\chi})U\right)^{-1} \right]^{mn}_{kl}
\end{align}
with $U^{ij}_{mn}$ being bare interaction parameters in the Wannier basis and $\tilde{\chi}$ the polarization restricted in the flat bands subspace using weighted method\cite{csacsiouglu2011effective}.
\begin{align}
\tilde{\chi}^{ij}_{kl}(\mathbf{q}, i \omega) &= \frac{1}{N_k} \sum_{n n' \mathbf{k}} \frac{f_{n \mathbf{k}}-f_{n' \mathbf{k}-\mathbf{q}}}{\epsilon_{n \mathbf{k}}-\epsilon_{n' \mathbf{k}-\mathbf{q}}-i \omega} \nonumber \\
&\quad \times p_{n \mathbf{k}} p_{n' \mathbf{k}-\mathbf{p}} u^*_{n \mathbf{k},i} u_{n' \mathbf{k}-\mathbf{q},j} u_{n' \mathbf{k}-\mathbf{q},l}^{*}u_{n' \mathbf{k},k} \\
U^{i j}_{ k l}(\mathbf{r}, \mathbf{r}') &= \iint d^3 r d^3 r' \left[ \varphi_i^*(\mathbf{r}) \varphi_j(\mathbf{r}) \right] V\left(\mathbf{r}, \mathbf{r}'\right) \left[ \varphi_k^*(\mathbf{r}') \varphi_l(\mathbf{r}') \right]
\end{align}
where $ p_{n \mathbf{k}}$ the weighted factor that account for how much the n-th band is within the target subspace, and $u_{n \mathbf{k},i}$ being the i-th component of the n-th band's cell periodic wavefunction, $\varphi_i(\mathbf{r})$ being i-th Wannier function.\\
The initial density–density interaction matrix elements were estimated using a preliminary cRPA calculation, as implemented in the VASP package \cite{Kresse-1993-VASP}. In this calculation, we employed a 24-orbital Wannier basis comprising 15 Co–3$d$ and 9 Sn–5$p$ orbitals as the target subspace. The exchange–correlation functional was treated within the PBE approximation \cite{Blochl-1994-PAW,Perdew-1996-PBE}, with an energy cutoff of 450 eV and a $8\times8\times9$ k-mesh. The resulting unscreened on-site interaction matrix elements are approximately $U^{ii}_{ii} \sim 23$ eV for the $d$ orbitals and 8.5 eV for the $p$ orbitals, while the inter-orbital terms yield $U^{ii}_{jj} \sim 21$ eV and 7.5 eV, respectively. Subsequent cRPA calculations performed on the lattice model give a screened on-site intra-orbital interaction of about 90 meV within the flat-band manifold.
However, it is crucial to note that calculations from such models are prone to over-screening effects\cite{shinaoka2015accuracy}. In comparison to established cRPA results where d-orbitals are excluded\cite{beinevig2025kagome_1}, this over-screening is expected to introduce an error on the order of unity. Therefore, the result from the lattice cRPA should be considered an order-of-magnitude estimate.

\end{document}